\documentclass[12pt]{iopart}

\expandafter\let\csname equation*\endcsname\relax
\expandafter\let\csname endequation*\endcsname\relax

\usepackage{amsmath,amssymb}
\usepackage{comment}
\usepackage{graphicx}
\usepackage{array}
\usepackage{color}

\begin{document}

\title{Non-Hermitian Floquet dynamics in absorption spectroscopy}

\author{R. M. Potvliege}

\address{Department of Physics, Durham University, South Road, Durham DH1 3LE, UK}
\ead{r.m.potvliege@durham.ac.uk}

\begin{abstract}
A theory of the absorption of a laser field by an atomic or condensed matter medium is presented for the case where the medium is also interacting with a strong electromagnetic field. The rotating wave approximation is not assumed for the latter. It is shown that in the weak probe limit the Lindblad master equation reduces to a smaller system of linear equations for the relevant steady state coherences. In this limit, the complex susceptibility of the medium can be expressed in terms of individual contributions of decaying dressed states, the latter being eigenstates of a non-Hermitian Floquet Hamiltonian.
\end{abstract}

\maketitle

\section{Introduction}

{\color{black}This article primarily concerns the theory of the absorption of a weak laser beam by an atomic or condensed matter medium also interacting with one or several other electromagnetic fields. Specifically, it focuses on the case where at least one of these other fields is too strong to be taken into account within the rotating wave approximation.} In the applications we have in mind, this strong field would typically be a microwave or radio frequency field; however, the theory is not limited to this particular case.
The relevance of the Floquet theory of linear differential equations with periodic coefficients in this context and in other areas of quantum optics has long been recognised  \cite{Shirley1965,Sambe1973,CohenTannoudji1977}.
{\color{black} Its relevance} is easily appreciated from the fact that the Hamiltonian describing the interaction of a quantum system with a CW monochromatic electromagnetic field is periodic in time, barring other time-dependent interactions. However, the theory also applies to stationary multi-frequency fields \cite{Ho1983,Chu2004,Joachain} and helps understand the dynamics of quantum systems interacting with non stationary fields, e.g., the dynamics of atoms in intense laser pulses \cite{Doerr1990,Day2000,Potvliege2006}.

A full Floquet description is often unnecessary for the relatively weak, near resonant fields typically used in spectroscopic measurements. The rotating wave approximation can usually be made for such fields, which simplifies the theory considerably \cite{Joachain,Foot}. However, making this approximation is not always {\color{black}justified}. For instance, going beyond the rotating wave approximation is normally essential in applications of Floquet theory to the modelling of multiphoton ionisation and other multiphoton processes driven by a strong laser field \cite{Chu2004,Joachain,Doerr1990,Potvliege2006,Chu1977,Potvliege1989}, 
to the measurement of low frequency fields using atoms in Rydberg states \cite{Anderson2014,Miller2016,Jiao2017,Paradis2019,Rotunno2023}, and of course for understanding how the properties of atomic or other systems can be manipulated through Floquet engineering --- e.g., \cite{Holthaus2015,Oka2019,Geier2021,Shan2021,Weitenberg2021,Yin2022,BlancoMas2023}.

Floquet calculations involves replacing the time-dependent Hamiltonian of the system of interest by a time-independent Hamiltonian acting in a larger Hilbert space, the ``Floquet Hamiltonian" \cite{Shirley1965,Sambe1973}. This Floquet Hamiltonian is non-Hermitian in some cases, for example in the cases considered in \cite{Wu2021} or in \cite{Zhou2023}. In particular, it is non-Hermitian in the complex dilation approach to multiphoton ionisation in a strong laser field \cite{Chu1977,Potvliege1989}: the radial variable $r$ is replaced by $r\exp(i\theta_{\rm r})$ for a well-chosen value of the angle $\theta_{\rm r}$, which transforms the Floquet Hamiltonian into a non-Hermitian operator. Doing so yields complex quasienergies, whose imaginary parts, within a factor $(-2/\hbar)$, give the ionisation rates of the states of interest \cite{Moiseyev1998}. This approach to multiphoton ionisation has been shown to be mathematically rigorous \cite{Howland1974,Yajima1982,Graffi1985}.
{\color{black}The Floquet Hamiltonian is also non-Hermitian for the systems we are studying in this work, but for a different reason.}

{\color{black}The present theory concerns the absorption of a weak beam by a medium modelled as an ensemble of individual systems with a finite number of bound energy eigenstates, for the case where some of these bound states may decay radiatively or otherwise.} 
For example, the methods outlined {\color{black} below} have recently been applied to the calculation of the absorption of a weak probe beam by a sample of cuprous oxide (Cu$_2$O) driven by a strong microwave field \cite{Brewin2024}. {\color{black} In this system, the probe beam couples broad Rydberg excitons decaying through interaction with phonons to the valence band (treated as a single bound state).} These methods could also be applied, e.g., to the spectroscopy of a strongly driven atomic vapour for which the excited atoms decay radiatively.
The quantum state of such media can usually be described by a density matrix satisfying the Lindblad master equation. Floquet approaches to this {\color{black}problem} have been discussed in the literature --- e.g., \cite{Ho1986,Jacobs2014,LeBoite2020,Chen2024} and, in regard to the optical properties of the medium, \cite{Wang1987,Nakano1994,Nakano1995,Sauvan2009,Tanasittikosol2012,Gu2018,Mishra2019}. {\color{black} However, and as we show in this article, a full Lindblad-Floquet calculation is not necessary in the (quite common) case where the probe field is weak enough to be treated in first order of perturbation theory. Indeed, the master equation reduces to a set of simpler equations in this case, with each decaying bound state being associated with a complex energy.} I.e., the real energies $\hbar\omega^{(i)}$, $i=0,1,2,\ldots$, of the different states included in the model are replaced by the complex energies $\hbar\omega^{(i)} - i\hbar\Gamma^{(i)}/2$, $i=0,1,2,\ldots$, where $\Gamma^{(i)}$ is the decay rate of state $i$. The properties of the medium can then be described in terms of the eigenstates of a non-Hermitian Floquet Hamiltonian, this Hamiltonian being non-Hermitian because of the non-zero imaginary parts of these complex energies.
In particular, the calculation yields the linear susceptibility of the medium as a sum of contributions from the decaying dressed states of the Floquet Hamiltonian. We stress that the imaginary parts of these energies are rigorously derived from the master equation, here, and are not {\it ad hoc} additional phenomenological decay widths.

For simplicity, we only consider the case of a homogeneous single species medium interacting with two monochromatic electromagnetic fields, namely a weak probe field and a strong coupling field, both polarised linearly in the same direction. The theory can be easily generalised to multi-species media and to more complex excitation pathways as long as the rotating wave approximation and the Beer-Lambert law apply for the probe field.

{\color{black} The general theory is developed in section~\ref{section:2} and its weak probe and long time limits in section~\ref{section:weakprobe}. A numerical illustration of the theory can be found in section~\ref{section:numerical}.}

\section{The probe absorption spectrum}
\label{section:2}

\subsection{The absorption coefficient}

As already mentioned, this article primarily concerns experiments in which a strong microwave or some other electromagnetic field (the ``coupling field") is applied to a homogeneous sample addressed by a laser beam (the ``probe field"), in circumstances where the latter is sufficiently weak that its attenuation is well described by Beer-Lambert law. We thus assume that
\begin{equation}
    I(x) = \exp(-K x) I(x=0),
\end{equation}
where $I(x)$ is the {\color{black} probe} intensity at a distance $x$ inside the medium and $K$, the absorption coefficient, is a constant.
For simplicity, we assume that {\color{black} the probe and coupling fields} are monochromatic and polarised in the $z$-direction, that the probe field can be modelled as a plane wave propagating in the $x$-direction and that the coupling field is homogeneous. We also assume that these fields are turned on at $t=0$. The details of the turn on are not important in so far as the state of the medium in the $t \rightarrow \infty$ limit is concerned, which is the focus of this work. Accordingly, we write their electric field components as follows,
\begin{align}
   {\bf E}_{\rm p}(x,t) &= \frac{1}{2}\, \widehat{\bf z} \,\left[ {\cal E}_{\rm p}(x)
\exp(-i \omega_{\rm p} t) + {\cal E}_{\rm p}^*(x)
\exp(i\omega_{\rm p} t)\right]H(t), \label{eq:Epdefined}\\
   {\bf E}_{\rm c}(t) &= \frac{1}{2}\, \widehat{\bf z} \,\left[ {\cal E}_{\rm c}
\exp(-i\omega_{\rm c} t) + {\cal E}_{\rm c}^*
\exp(i\omega_{\rm c} t)\right]H(t),
\label{eq:Ecdefined}
\end{align}
where $\widehat{\bf z}$ is a unit vector in the $z$-direction and $H(t)$ is the Heaviside step function.
The probe and coupling fields polarise the medium, resulting in a polarisation field ${\bf P}(x,t)$.
In terms of the density operator 
$\hat{\rho}(x,t)$ describing the state of the medium at position $x$ and time $t$, 
\begin{equation}
    {\bf P}(x,t) = N_{\rm d}\,\mbox{Tr}[\,\hat{\rho}(x,t)\,\hat{D}_z\,]\,\widehat{\bf z},
\label{eq:Pfromrho}
\end{equation}
where $\hat{D}_z$ is the $z$-component of the dipole operator and $N_{\rm d}$ is the number density of atoms or other individual systems whose state is described by $\hat{\rho}(x,t)$.
Assuming that $t$ is large enough for the medium to have evolved into a stationary state,
\begin{align}
   {\bf P}(x,t) &= \frac{1}{2}\, \widehat{\bf z} \,\left[ {\cal P}_{\rm p}(x)
\exp(-i\omega_{\rm p} t) + {\cal P}_{\rm p}^*(x)
\exp(i\omega_{\rm p} t) + \ldots\right],
\label{eq:Pdefined}
\end{align}
where $\ldots$ stands for contributions oscillating at angular frequencies other than $\omega_{\rm p}$. The complex polarisation amplitude ${\cal P}_{\rm p}(x)$ is proportional to ${\cal E}_{\rm p}(x)$ for the weak probe fields considered here: in terms of a frequency-dependent susceptibility $\chi(\omega_{\rm p},\omega_{\rm c},{\cal E}_{\rm c})$,
\begin{align}
{\cal P}_{\rm p}(x) &= {\epsilon_0}\,\chi(\omega_{\rm p},\omega_{\rm c},{\cal E}_{\rm c})\,{\cal E}_{\rm p}(x).
\label{eq:chidefined}
\end{align}
If follows from the above \cite{Loudon} that
\begin{equation}
    K = 2 (\omega_{\rm p}/c)\,\mbox{Im}\,[1 + \chi(\omega_{\rm p},\omega_{\rm c},{\cal E}_{\rm c})]^{1/2}.
    \label{eq:Kcoef}
\end{equation}
{\color{black} The absorption coefficient $K$ does not depend on $x$ since the coupling field is taken to be homogeneous.} Accordingly, we no longer specify the spatial dependence of ${\cal E}_{\rm p}(x)$, ${\cal P}_{\rm p}(x)$ and $\hat{\rho}(x,t)$.

\subsection{Microscopic description}

We use a finite basis of ${\cal N}$ orthonormal eigenstates of the medium's field-free Hamiltonian, $\hat{H}_0$, namely a basis $\{|i\rangle, i=0,\ldots,{\cal N}-1\}$ such that $\langle i | j \rangle = \delta_{ij}$ and
\begin{equation}
\hat{H}_0 |i\rangle= \hbar \omega^{(i)} |i\rangle, \qquad i = 0,\ldots,{\cal N}-1.
\label{eq:H0}
\end{equation}
We describe the interaction between the medium and the applied fields within the electric dipole approximation. The Hamiltonian of the system can therefore be written in the following way in terms of the Rabi frequencies
    ${\Omega}_{{\rm p};ij} = 
{\cal E}_{\rm p}\,\langle i |
 \hat{D}_z | j \rangle/\hbar$
and ${\Omega}_{{\rm c};ij} = 
{\cal E}_{\rm c}\,\langle i |
 \hat{D}_z | j \rangle/\hbar$:
\begin{align}
    \hat{H}(t) = \sum_{i} \hbar \omega^{(i)} |i\rangle\langle i| - \frac{\hbar}{2}\sum_{i,j} &\left\{\left[
    {\Omega}_{{\rm p};ij} \exp(-i\omega_{\rm p} t) +
    {\Omega}_{{\rm c};ij} \exp(-i\omega_{\rm c} t)
     \right] |i\rangle\langle j| + \mbox{h.c.}\right\}.
    \label{eq:fullH}
\end{align}

Typically,
the relevant states form two groups differing considerably in energy, namely a low energy group~${}A$ and a higher energy group~${}B$. We will assume that the coupling field strongly couples the states of  group~${}B$ to each other, but does not directly couple the states of group~${}A$ to each other or to those of group~${}B$, while the probe field only couples states of group~${}A$ to states of group~${}B$. Neglecting the far-detuned transitions, we thus set
\begin{align}
    \hat{H}(t) = \sum_{i} \hbar \omega^{(i)} |i\rangle\langle i| &- \frac{\hbar}{2}\sum_{i\in{}A}\sum_{j\in {}B} \left[
    {\Omega}_{{\rm p};ij} \exp(-i\omega_{\rm p} t)
      |i\rangle\langle j| + \mbox{h.c.}\right] \nonumber \\
 &- \frac{\hbar}{2}\sum_{i\in{}B}\sum_{j\in {}A} \left[
    {\Omega}_{{\rm p};ij} \exp(-i\omega_{\rm p} t)
      |i\rangle\langle j| + \mbox{h.c.}\right] \nonumber \\
     &- \frac{\hbar}{2}\sum_{i\in{}B}\sum_{j\in {}B} \left[
    {\Omega}_{{\rm c};ij} \exp(-i\omega_{\rm c} t)
      |i\rangle\langle j| + \mbox{h.c.}\right].
    \label{eq:simplifiedH}
\end{align}
Given this Hamiltonian, the density operator $\hat{\rho}(t)$ is governed by the Lindblad master equation
\begin{equation}
\frac{{\rm d} \hat{\rho}}{{\rm d} t} =
-\frac{i}{\hbar}\,[\hat{H},\hat{\rho}\,] + 
\frac{1}{2} \sum_{n} \left(
2\, \hat{C}_n \hat{\rho} \, \hat{C}_n^\dagger - \hat{C}_n^\dagger \hat{C}_n \hat{\rho} - \hat{\rho}\, \hat{C}_n^\dagger \hat{C}_n
\right),
\label{eq:Lindblad}
\end{equation}
where the $\hat{C}_n$'s are the collapse (or jump) operators accounting for decoherence and relaxation. The  $\hat{C}_n$'s should
include the operator 
$\sqrt{\Gamma_{ij}} \, |i\rangle \langle j|$ if state $j$ relaxes to state $i$ at a rate $\Gamma_{ij}$, e.g., by spontaneous decay.

\subsection{The rotating wave approximation}

We will also assume that the energies are measured with respect to a given reference energy $\hbar \omega^{\rm ref}$ chosen so that the differences $|\omega^{(i)}-\omega^{\rm ref}|$ are much smaller for the states of group~${}A$ than for those of group~${}B$. Conversely, the differences $|\omega_{\rm p} - (\omega^{(i)}-\omega^{\rm ref})|$ are much smaller for the states of group~${}B$ than for those of group~${}A$. We can thus pass to slowly varying variables by transforming state vectors and density operators by the unitary transformation\footnote{\color{black}This transformation is unitary in the Hilbert space spanned by the basis of ${\cal N}$ eigenvectors of the field-free Hamiltonian used throughout these calculations: $\hat{U}(t)\hat{U}^\dagger(t) = \hat{U}^\dagger(t)\hat{U}(t) = \sum_{i\in {}A}| i \rangle \langle i| + \sum_{j\in {}B}| j \rangle \langle j|$, which is the identity operator in that Hilbert space.}
\begin{align}
        \hat{U}(t) = \exp\left(i\omega^{\rm ref}t\right) \sum_{i\in {}A}| i \rangle \langle i| &+
        \exp\left[i\left(\omega^{\rm ref} +\omega_{\rm p}\right)t\right] 
        \sum_{j\in {}B}| j \rangle \langle j|.
    \label{eq:Ucase4}
\end{align}
In particular, the operator $\hat{U}(t)$ transforms $\hat{\rho}(t)$ into the density operator 
\begin{equation}
\hat{\rho}^{\rm tr}(t) = \hat{U}(t)\hat{\rho}(t)\hat{U}^\dagger(t),
\end{equation}
and
\begin{equation}
    \langle i | \hat{\rho}(t) | j\rangle =
    \begin{cases}
        \langle i | \hat{\rho}^{\rm tr}(t) | j \rangle & \mbox{if $i$ and $j \in {}A$}, \\
        \exp(i\omega_{\rm p} t) \langle i | \hat{\rho}^{\rm tr}(t) | j \rangle & \mbox{if $i \in {}A$ and $j \in {}B$}, \\
        \exp(-i\omega_{\rm p} t) \langle i | \hat{\rho}^{\rm tr}(t) | j \rangle & \mbox{if $i \in {}B$ and $j \in {}A$}, \\
        \langle i | \hat{\rho}^{\rm tr}(t) | j \rangle & \mbox{if $i$ and $j \in {}B$}.
    \end{cases}
\end{equation}
A short calculation shows that $\hat{\rho}^{\rm tr}(t)$ evolves in time according to the equation
\begin{equation}
\frac{{\rm d} \hat{\rho}^{\rm tr}}{{\rm d} t} =
-\frac{i}{\hbar}\,[\hat{H}^{\rm tr},\hat{\rho}^{\rm tr}\,] + 
\frac{1}{2} \sum_{n} \left(
2\, \hat{C}_n \hat{\rho}^{\rm tr} \hat{C}_n^\dagger - \hat{C}_n^\dagger \hat{C}_n \hat{\rho}^{\rm tr} - \hat{\rho}^{\rm tr} \hat{C}_n^\dagger \hat{C}_n
\right),
\label{eq:Lindblad2}
\end{equation}
where
\begin{align}
    \hat{H}^{\rm tr}(t) = \hbar \sum_{i \in {}A} \delta\omega^{(i)} |i\rangle\langle i| &
    -\hbar \sum_{i \in {}B} \Delta_{\rm p}^{(i)} |i\rangle\langle i| 
    - \frac{\hbar}{2}\sum_{i\in{}B}\sum_{j\in {}A} \left(
    {\Omega}_{{\rm p};ij}
     |i\rangle\langle j| + \mbox{h.c.}\right) \nonumber \\
 &- \frac{\hbar}{2}\sum_{i\in{}A}\sum_{j\in {}B} \left[
    {\Omega}_{{\rm p};ij} \exp(-2i\omega_{\rm p} t)
     |i\rangle\langle j| + \mbox{h.c.}\right] \nonumber \\
     &- \frac{\hbar}{2}\sum_{i\in{}B}\sum_{j\in {}B} \left[
    {\Omega}_{{\rm c};ij} \exp(-i\omega_{\rm c} t)
      |i\rangle\langle j| + \mbox{h.c.}\right]
    \label{eq:Htransformed}
\end{align}
with $\delta \omega^{(i)}=\omega^{(i)}-\omega^{\rm ref}$ and $\Delta_{\rm p}^{(i)} = \omega_{\rm p}-(\omega^{(i)}-\omega^{\rm ref})$. 

{\color{black} We now make the rotating wave approximation for the probe field}, which is to neglect the rapidly oscillating terms in $\exp(\pm 2 i \omega_{\rm p}t)$ appearing in equation~(\ref{eq:Htransformed}). Accordingly, we set
\begin{equation}
    \hat{\rho}(t) = \hat{U}^\dagger(t)\hat{\rho}^{\rm rw}(t)\hat{U}(t)
\label{eq:rhorwdefined}
\end{equation}
where $\hat{\rho}^{\rm rw}(t)$ satisfies the equation
\begin{equation}
\frac{{\rm d} \hat{\rho}^{\rm rw}}{{\rm d} t} =
-\frac{i}{\hbar}\,[\hat{H}^{\rm rw},\hat{\rho}^{\rm rw}\,] + 
\frac{1}{2} \sum_{n} \left(
2\, \hat{C}_n \hat{\rho}^{\rm rw}  \hat{C}_n^\dagger - \hat{C}_n^\dagger \hat{C}_n \hat{\rho}^{\rm rw} - \hat{\rho}^{\rm rw} \hat{C}_n^\dagger \hat{C}_n
\right)
\label{eq:Lindblad3}
\end{equation}
with
\begin{align}
    \hat{H}^{\rm rw}(t) = \hbar \sum_{i \in {}A} \delta\omega^{(i)} |i\rangle\langle i| &
    -\hbar \sum_{i \in {}B} \Delta_{\rm p}^{(i)} |i\rangle\langle i| 
    - \frac{\hbar}{2}\sum_{i\in{}B}\sum_{j\in {}A} \left(
    {\Omega}_{{\rm p};ij}
     |i\rangle\langle j| + \mbox{h.c.}\right) \nonumber \\
     &- \frac{\hbar}{2}\sum_{i\in{}B}\sum_{j\in {}B} \left[
    {\Omega}_{{\rm c};ij} \exp(-i\omega_{\rm c} t)
      |i\rangle\langle j| + \mbox{h.c.}\right].
    \label{eq:Htransformedrw}
\end{align}
The density operator $\hat{\rho}^{\rm rw}(t)$ is represented by the matrix $[\rho_{ij}^{\rm rw}(t)]$ in the basis $\{|0\rangle,\ldots,|{\cal N}-1\rangle\}$, with
\begin{equation}
\rho_{ij}^{\rm rw}(t) = \langle i | \hat{\rho}^{\rm rw}(t)| j \rangle.
\end{equation}
Arranging the matrix elements $\rho_{ij}^{\rm rw}(t)$ into a column vector ${\sf r}(t)$ makes it possible to recast equation~(\ref{eq:Lindblad3}) into a more standard form, namely
\begin{equation}
    \frac{{\rm d}{\sf r}}{{\rm d}t} = {\sf L}(t) {\sf r}(t),
\label{eq:rdot}
\end{equation}
where ${\sf L}(t)$ is a ${\cal N}^2 \times {\cal N}^2$ matrix.

\subsection{Floquet formulation}
{\color{black} 
In principle, calculating the absorption coefficient reduces to solving equation~(\ref{eq:Lindblad3}) or equation (\ref{eq:rdot}), calculating the resulting polarisation field ${\bf P}(x,t)$ and obtaining the polarisation amplitude ${\cal P}_{\rm p}(x)$ by Fourier-transforming ${\bf P}(x,t)$. As will become clear later on, however, there are advantages to first calculating the Fourier components of the density operator, and thereby seek solutions of the Lindblad master equation in the Floquet form.
We note, in this regard,} that ${\sf L}(t)$ can be written in terms of three constant matrices, ${\sf L}_0$, ${\sf L}_+$ and ${\sf L}_-$, in the following way:
\begin{equation}
    {\sf L}(t) = {\sf L}_0 + \exp(i\omega_{\rm c} t){\sf L}_- + \exp(-i\omega_{\rm c}t){\sf L}_+.
\end{equation}
The time-periodicity of ${\sf L}(t)$ suggests to seek a solution of equation~(\ref{eq:rdot}) of the form
\begin{equation}
    {\sf r}(t) = \sum_{N} \exp(-i N \omega_{\rm c} t) {\sf r}_N(t),
\label{eq:rexpanded}
\end{equation}
the ${\sf r}_N(t)$ being time-dependent column vectors such that
\begin{equation}
   \frac{{\rm d}{\sf r}_N}{{\rm d}t} = {\sf A}_N{\sf r}_N(t) + {\sf L}_+{\sf r}_{N-1}(t) + {\sf L}_-{\sf r}_{N+1}(t), \quad N = 0,\pm 1, \pm 2,\ldots,
\label{eq:drndt}
\end{equation}
where ${\sf A}_N = {\sf L}_0 + iN\omega_{\rm c}{\sf I}$ with ${\sf I}$ denoting the ${\cal N}^2 \times {\cal N}^2$ unit matrix. In practice, this infinite system can be truncated to a finite number of equations and the index $N$ restricted to a certain interval $[N_{\rm min},N_{\rm max}]$.
equation~(\ref{eq:drndt}) can be expressed in matrix form as
\begin{align}
\frac{{\rm d}\;}{{\rm d}t}
\begin{pmatrix}
    \vdots \\ {\sf r}_{-2}(t) \\ {\sf r}_{-1}(t) \\ {\sf r}_0(t) \\ {\sf r}_{1}(t)  \\ {\sf r}_{2}(t)\\ \vdots
    \end{pmatrix}
    = &\begin{pmatrix}
    &&& \vdots\\
    {\sf L}_+ & {\sf A}_{-2} & {\sf L}_- &&&&\\
    &{\sf L}_+ & {\sf A}_{-1} & {\sf L}_- &&&\\
    &&{\sf L}_+ & {\sf A}_0 & {\sf L}_- && \\
    &&& {\sf L}_+ & {\sf A}_{1} & {\sf L}_- &\\
    &&&& {\sf L}_+ & {\sf A}_{2} & {\sf L}_- \\
    &&& \vdots
    \end{pmatrix}\begin{pmatrix}
    \vdots \\ {\sf r}_{-2}(t) \\ {\sf r}_{-1}(t) \\ {\sf r}_0(t) \\ {\sf r}_{1}(t)  \\ {\sf r}_{2}(t)\\ \vdots
    \end{pmatrix},
\end{align}
or, in a more compact notation, as
\begin{equation}
\frac{{\rm d}{\sf R}}{{\rm d}t} = {\sf M}{\sf R}(t).
\label{eq:drndtmatrix}
\end{equation}
{\color{black}
Apart from the truncation of the Fourier expansion to a finite number of harmonic components, calculating the density matrix through this last equation and equation~(\ref{eq:rexpanded}) is equivalent to solving the Lindblad equation~(\ref{eq:Lindblad3}). These two approaches produce numerically different results for finite values of $N_{\rm min}$ and $N_{\rm max}$; however, and as illustrated by an example in Section~\ref{section:numerical}, these differences can be made arbitrarily small by taking the interval $[N_{\rm min},N_{\rm max}]$ wide enough.
}

The density matrix describing the state of the medium can therefore be obtained in terms of the eigenvalues and eigenvectors of the non-Hermitian constant matrix ${\sf M}$, unless this matrix would be defective. Let us denote the eigenvalues of ${\sf M}$ by $w^{(k)}$, its right eigenvectors by ${\sf c}^{(k)}$ and its left eigenvectors by ${\sf d}^{(k)}$, so that ${\sf M}{\sf c}^{(k)} = w^{(k)} {\sf c}^{(k)}$ and ${\sf d}^{(k)\dagger}{\sf M} = w^{(k)} {\sf d}^{(k)\dagger}$ (the ${\sf c}^{(k)}$'s and the ${\sf d}^{(k)}$'s are column vectors). Assuming that these eigenvectors form a basis and are bi-orthogonal (i.e., that ${\sf d}^{(k)\dagger}{\sf c}^{(k')} = 0$ if $k\not= k'$),
\begin{equation}
    {\sf R}(t) = \sum_k 
    \alpha_k {\sf c}^{(k)} \exp\left(w^{(k)} t\right)
\end{equation}
with 
\begin{equation}
    \alpha_k = \frac{{\sf d}^{(k)\dagger}{\sf R}(t=0)}{{\sf d}^{(k)\dagger}{\sf c}^{(k)}}.
\end{equation}
It is worth noting that the right eigenvectors of ${\sf M}$ define solutions of equation~(\ref{eq:rdot}) in the Floquet form, namely
\begin{equation}
    {\sf r}(t) = \exp\left(w^{(k)} t\right)\sum_N \exp(-iN\omega_{\rm c}t){\sf c}^{(k)}_{N}.
\end{equation}
The eigenvectors for which $w^{(k)} = 0$ may describe steady states of the system, i.e., states for which ${\rm d}{\sf r}_N/{\rm d}t = 0$. However, the solutions with $w^{(k)} \not =0$ do not correspond to density operators of constant unit trace and therefore do not represent physically meaningful quantum states.

Since the column vector ${\sf r}(t)$ is formed by the matrix elements $\rho_{ij}^{\rm rw}(t)$, the elements of the vectors ${\sf r}_N(t)$ are the functions $\rho_{ij;N}^{\rm rw}(t)$ such that
\begin{equation}
    \rho_{ij}^{\rm rw}(t) = \sum_{N} \exp(-iN\omega_{\rm c}t)\rho_{ij;N}^{\rm rw}(t).
\label{eq:rhoijexp}
\end{equation}
As will be illustrated by a numerical example in Section~\ref{section:numerical}, the functions $\rho_{ij;N}^{\rm rw}(t)$ normally converge to constant values in the $t \rightarrow \infty$ limit. In the ensuing steady state, the density matrix contains terms oscillating at the angular frequencies $\omega_{\rm p}$, $\omega_{\rm p} \pm \omega_{\rm c}$, $\omega_{\rm p} \pm 2 \omega_{\rm c}$, etc. Only the terms oscillating at the angular frequency $\omega_{\rm p}$ contribute to the complex polarisation amplitude ${\cal P}_{\rm p}(x)$. Combining equations~(\ref{eq:Pfromrho}), (\ref{eq:Pdefined}), (\ref{eq:chidefined}), (\ref{eq:Ucase4}), (\ref{eq:rhorwdefined}) and (\ref{eq:rhoijexp}) yields
\begin{equation}
    \chi(\omega_{\rm p},\omega_{\rm c},{{\cal E}_{\rm c}}) = \frac{N_{\rm d}}{\epsilon_0{\cal E_{\rm p}}} \sum_{i\in {}B} \sum_{j\in {}A} \langle j|\hat{D}_z|i\rangle \rho_{ij;N=0}^{\rm rw}(t \rightarrow \infty).
\label{eq:chi}
\end{equation}
Having calculated $\chi(\omega_{\rm p},\omega_{\rm c},{{\cal E}_{\rm c}})$, the probe absorption spectrum can then be deduced from equation~(\ref{eq:Kcoef}).

{\color{black}It is worth noting that the advantage of the Floquet approach, in this context, is to yield the $N=0$ harmonic component of the relevant coherences directly. Merely calculating time-dependent populations and coherences is likely to be more conveniently done by solving equation~(\ref{eq:Lindblad3}) than by solving equation~(\ref{eq:drndtmatrix}). The Floquet approach has marked advantages for calculations in the weak probe limit, however, as we now discuss.
}

\section{The weak probe limit}
\label{section:weakprobe}
\subsection{General theory}
\label{section:weakgeneral}

It is often the case, in experiments, that the medium can be taken to be initially in its ground state or in an incoherent superposition of low energy states, and that the probe laser is too weak for producing significant optical pumping over the relevant time scales. This common situation can be modelled by assuming, (1) that the probe field is weak enough that $\hat{\rho}^{\rm rw}(t)$ only needs to be calculated to first order in ${\cal E}_{\rm p}$, and (2) that the states of group~${}B$ are initially unpopulated. {\color{black} The latter implies that $\rho_{ij}^{\rm rw}(t) = 0$ at $t=0$ if state~$i$ or state~$j$ belongs to group~${}B$ and that
\begin{equation}
    \sum_{i \in {}A} \rho_{ii}^{\rm rw}(t=0) = 1.
    \end{equation}
}
{\color{black}We will further assume that $\rho_{ij}^{\rm rw}(t=0)=0$ for $j\not= i$ and $\Gamma_{ij} = 0$} if states $i$ and $j$ both belong to group~${}A$, which is normally the case in applications. The density operator $\hat{\rho}^{\rm rw}(t)$ reduces to a density operator $\hat{\rho}^{\rm wp}(t)$ in this weak probe approximation:
\begin{equation}
{\rho}_{ij}^{\rm rw}(t)\approx \rho_{ij}^{\rm wp}(t)
\end{equation} 
with ${\rho}^{\rm wp}_{ij}(t=0) = {\rho}^{\rm rw}_{ij}(t=0)$.
One finds, after a straightforward calculation, that 
\begin{align}
    \frac{{\rm d}\rho_{ij}^{\rm wp}}{{\rm d}t} = 0
\end{align}
if state $i$ and state $j$ both belong to group~${}A$ or both belong to group~${}B$, and that
\begin{align}
    \frac{{\rm d}\rho_{ij}^{\rm wp}}{{\rm d}t} &= i[\omega^{(j)} + \omega_{\rm p} - \omega^{(i)} + i \Gamma^{(i)}/2 ] \rho_{ij}^{\rm wp}(t) + \frac{i}{2}
    \,\Omega_{{\rm p};ij}\rho_{jj}^{\rm wp}(t)
    \nonumber \\ &\qquad \qquad +\frac{i}{2} \sum_{l\in {}B} [\Omega_{{\rm c};il}\exp(-i\omega_{\rm c}t)+\Omega_{{\rm c};li}^*\exp(i\omega_{\rm c}t)] \rho_{lj}^{\rm wp}(t)
\label{eq:drhowp}
\end{align}
if $i\in {}B$ and $j \in {}A$. In this last equation, $\Gamma^{(i)}$ denotes the total decay rate of state~$i$:
\begin{equation}
    \Gamma^{(i)} = \sum_n \Gamma_{ni}.
\end{equation}
{\color{black} In view of equation~(\ref{eq:chi}), our main aim is to solve equation~(\ref{eq:drhowp}) in the long time limit.} Proceeding as above, we seek a solution of equation~(\ref{eq:drhowp}) of the form
\begin{equation}
    \rho_{ij}^{\rm wp}(t) = \sum_{N} \exp(-iN\omega_{\rm c}t)\rho_{ij;N}^{\rm wp}(t), 
\label{eq:rhoijwp}
\end{equation}
with
\begin{align}
    \frac{{\rm d}\rho_{ij;N}^{\rm wp}}{{\rm d}t} &= i[\omega^{(j)} + \omega_{\rm p} - \omega^{(i)} + i \Gamma^{(i)}/2 + N\omega_{\rm c} ] \rho_{ij;N}^{\rm wp}(t) + \frac{i}{2}
    \,\Omega_{{\rm p};ij}\rho_{jj;N}^{\rm wp}(t)
    \nonumber \\ &\qquad \qquad +\frac{i}{2} \sum_{l\in {}B} [\Omega_{{\rm c};il}\rho_{lj;N-1}^{\rm wp}(t)+\Omega_{{\rm c};li}^* \rho_{lj;N+1}^{\rm wp}(t)].
\label{eq:drhowpharmonic}
\end{align}
Since the populations of the different states remain constant in this weak probe approximation,
\begin{equation}
    \rho_{jj;N}^{\rm wp}(t) \equiv  \rho_{jj;0}^{\rm wp}\, \delta_{N,0} = \rho_{jj}^{\rm wp}(t=0). 
\end{equation}
The steady state solution of equation~(\ref{eq:drhowpharmonic})
can be obtained by setting the time derivatives to zero and solving the resultant system of linear equations, namely
\begin{align}
    [\omega^{(j)} + \omega_{\rm p} &- \omega^{(i)} + i \Gamma^{(i)}/2 + N\omega_{\rm c} ] \rho_{ij;N}^{\rm wp}
    \nonumber \\ &\qquad  +\frac{1}{2} \sum_{l\in {}B} [\Omega_{{\rm c};il}\rho_{lj;N-1}^{\rm wp}+\Omega_{{\rm c};li}^* \rho_{lj;N+1}^{\rm wp}] = -\frac{1}{2}
    \,\Omega_{{\rm p};ij}\rho_{jj;0}^{\rm wp}\delta_{N,0},
\label{eq:steadysystem}
\end{align}
where $i \in {}B$, $j \in {}A$, $N_{\rm min}\leq N\leq N_{\rm max}$ and $\rho_{ij;N}^{\rm wp} \equiv \rho_{ij;N}^{\rm wp}(t\rightarrow \infty)$. Calculating the absorption spectrum in the weak probe limit thus reduces to solving this system of equations for each state of group~${}A$ and obtaining the absorption coefficient from equation~(\ref{eq:chi}) with $\rho_{ij;0}^{\rm rw}(t\rightarrow\infty)$ replaced by $\rho_{ij;0}^{\rm wp}$ --- i.e., from the equation
\begin{equation}
    \chi(\omega_{\rm p},\omega_{\rm c},{{\cal E}_{\rm c}}) = \frac{N_{\rm d}}{\epsilon_0{\cal E_{\rm p}}} \sum_{i\in {}B} \sum_{j\in {}A} \langle j|\hat{D}_z|i\rangle \rho_{ij;N=0}^{\rm wp}.
\label{eq:chiwp}
\end{equation}
Because $\Omega_{{\rm p};ij}\propto {\cal E}_{\rm p}$, $\rho_{ij;N=0}^{\rm wp} \propto {\cal E}_{\rm p}$ and the susceptibility does not depend on ${\cal E}_{\rm p}$ in this approximation. 

We note that only the total dephasing rates $\Gamma^{(i)}/2$ appear in this formulation of the optical Bloch equations, rather than the individual relaxation rates $\Gamma_{ij}$.
Contrary to equation~(\ref{eq:Lindblad2}),
equation~(\ref{eq:steadysystem}) therefore applies even to cases where the states of group~${}B$ decay to other states than those of group~${}A$, which makes it well suited for describing systems with complex de-excitation pathways.

We also note that the equations forming this system decouple from each other in the absence of the coupling field (i.e., for ${\cal E}_{\rm c} = 0$); equation~(\ref{eq:steadysystem}) then yields
\begin{equation}
    \rho_{ij;0}^{\rm wp}
     = -\frac{
    \Omega_{{\rm p};ij}/2}{\omega^{(j)} + \omega_{\rm p} - \omega^{(i)} + i \Gamma^{(i)}/2}\,\rho_{jj;0}^{\rm wp},
\label{eq:zerocfield}
\end{equation}
in agreement with well known theory --- see, e.g., \cite{Zentile2015}.

\subsection{Complex energies}
\label{section:complexenergies}

Equation~(\ref{eq:steadysystem}) was derived from the Lindblad master equation in the above, starting from the Hermitian Hamiltonian $\hat{H}(t)$ of equation~(\ref{eq:fullH}). However, it is easy to see that the same result would also be obtained, in the weak probe approximation,
if the density operator was instead calculated by solving the equation
\begin{equation}
\frac{{\rm d} \hat{\rho}}{{\rm d} t} =
-\frac{i}{\hbar}\,[\hat{H}',\hat{\rho}\,]
\label{eq:vonN}
\end{equation}
with $\hat{H}'(t)$ taken to be the non-Hermitian Hamiltonian
\begin{align}
    \hat{H}'(t) = \sum_{i \in {}A} & \hbar \omega^{(i)} |i\rangle\langle i| + \sum_{i \in {}B} \hbar \left(\omega^{(i)} - i  \Gamma^{(i)}/2\right) |i\rangle\langle i| \nonumber \\
    & \qquad - \frac{\hbar}{2}\sum_{i,j} \left\{\left[
    {\Omega}_{{\rm p};ij} \exp(-i\omega_{\rm p} t) +
    {\Omega}_{{\rm c};ij} \exp(-i\omega_{\rm c} t)
     \right] |i\rangle\langle j| + \mbox{h.c.}\right\}.
    \label{eq:fullH2}
\end{align}
In this alternative approach, the states of group~${}B$ are given a phenomenological width $\hbar \Gamma^{(i)}$ and the steady state of the system is derived solely from the effective Hamiltonian $\hat{H}'(t)$ by integrating the von~Neumann equation. The two approaches are equivalent in the weak probe approximation in so far as they predict the same results. However, they would generally lead to different results beyond that approximation, in which case energy relaxation needs to be taken into account through the Lindblad master equation rather than through complex energies.

\subsection{Dressed states}
\label{section:dressedstates}

Equation~(\ref{eq:steadysystem}) has a close similarity with the equation defining the dressed states of the system in the absence of the probe field, as we now discuss. By dressed states, we mean, here, the quasistationary solutions of the equation
\begin{equation}
\left[i\hbar\frac{{\rm d}\;}{{\rm d}t} - \sum_{i \in {}B} \hbar \left(\omega^{(i)} - i \Gamma^{(i)}/2\right)|i\rangle\langle i| +
{\bf E}_{\rm c}(t)\cdot\hat{\bf D}\right]
|\Psi(t)\rangle = 0,
\end{equation}
where $\hat{\bf D}$ is the dipole operator. The dressed states in question are solutions of the Floquet form, 
\begin{equation}
    |\Psi^{(q)}(t)\rangle = \exp(-i \epsilon^{(q)} t/\hbar)     |\Phi^{(q)}(t)\rangle,
\end{equation}
where $ \epsilon^{(q)}$ is a constant and the state vector
$|\Phi^{(q)}(t)\rangle$ is time-periodic with period $2\pi/\omega_{\rm c}$. We expand the latter in a Fourier series with time-independent harmonic components $|\psi^{(q)}_N\rangle$ and write
\begin{equation}
    |\Psi^{(q)}(t)\rangle =
    \exp(-i \epsilon^{(q)} t/\hbar) \sum_N \exp(-iN\omega_{\rm c}t)|\psi^{(q)}_N\rangle.
\end{equation}
Since the field ${\bf E}_c(t)$ only couples states belonging to group~${}B$ to other states of that group, we set
\begin{equation}
|\psi^{(q)}_{N}\rangle = \sum_{i \in {}B} a^{(q)}_{i;N} |i\rangle.
\end{equation}
It is not difficult to show that 
the coefficients $a^{(q)}_{i;N}$ satisfy a system of equations similar (but not identical) to equation~(\ref{eq:steadysystem}), i.e.,
\begin{align}
    \left[\epsilon^{(q)}- \hbar\left(\omega^{(i)} - i \Gamma^{(i)}/2\right) + N\hbar\omega_{\rm c}\right] a^{(q)}_{i;N}
    +\frac{\hbar}{2} \sum_{l\in {}B} \left[\Omega_{{\rm c};il}a^{(q)}_{l;N-1}+\Omega_{{\rm c};li}^* a^{(q)}_{l;N+1}\right] = 0.
\label{eq:Floquet}
\end{align}
In matrix form,
\begin{equation}
\left[\epsilon^{(q)}\,{\sf I}_{\rm Fl} - {\sf F} \right]{\sf v}^{(q)} = 0,
\label{eq:Feq0}
\end{equation}
where ${\sf v}^{(q)}$ is the column vector formed by the coefficients $a_{i;N}^{(q)}$, ${\sf I}_{\rm Fl}$ is the ${\cal N}_{\rm Fl}\times {\cal N}_{\rm Fl}$ unit matrix with ${\cal N}_{\rm Fl} = (N_{\rm max}-N_{\rm min}+1){\cal N}_B$, where ${\cal N}_B$ is the number of states belonging to group~${}B$, and ${\sf F}$ is a ${\cal N}_{\rm Fl}\times {\cal N}_{\rm Fl}$ non-Hermitian matrix. This matrix represents the Floquet Hamiltonian and describes how the states of group~${}B$ are coupled to each other by the coupling field beyond the rotating wave approximation.

How these dressed states relate to the solutions of equation~(\ref{eq:steadysystem}) may be best brought out by recasting the latter in terms of the matrix ${\sf F}$. For each state~$j \in A$, we write equation~(\ref{eq:steadysystem}) as
\begin{equation}
\left[(\omega^{(j)} + \omega_{\rm p})\,{\sf I}_{\rm Fl} - {\sf F}/\hbar \right]{\sf x}^{(j)} = {\sf b}^{(j)},
\label{eq:Feq}
\end{equation}
where ${\sf I}_{\rm Fl}$ and ${\sf F}$ are the matrices defined above (the same for all states $j$), ${\sf x}^{(j)}$ is the column vector formed by the coherences $\rho_{ij;N}^{\rm wp}$ and ${\sf b}^{(j)}$ is the column vector formed by the right-hand sides of equation~(\ref{eq:steadysystem}).
We will assume that ${\sf F}$ is not defective in view of the fact that this matrix would be Hermitian if the decay rates $\Gamma_i$ were zero. The results outlined below do not apply if ${\sf F}$ is defective. The coherences derived from Eq.~(\ref{eq:xexpanded2}) would be incorrect and would differ from those obtained by solving Eq.~(\ref{eq:steadysystem}) in this case. 

Clearly, the eigenvalues of ${\sf F}$ are the complex quasienergies $\epsilon^{(q)}$ defined above and its right eigenvectors are the column vectors ${\sf v}^{(q)}$:
\begin{equation}
    {\sf F}\,{\sf v}^{(q)} = \epsilon^{(q)}\,{\sf v}^{(q)}.
\label{eq:Floqueteigenvalr}
\end{equation}
We denote the corresponding left eigenvectors by ${\sf u}^{(q)}$ and define them as being the column vectors such that
\begin{equation}
    {\sf u}^{(q)\dagger}\,{\sf F} = \epsilon^{(q)}\,{\sf u}^{(q)\dagger}.
\label{eq:Floqueteigenval}
\end{equation}
Let ${\sf U}$ and ${\sf V}$ be the matrices formed by the column vectors ${\sf u}^{(q)}$ and ${\sf v}^{(q)}$, respectively, with ${\sf u}^{(q)}$ and ${\sf v}^{(q)}$ normalised in such a way that ${\sf u}^{(q)\dagger}{\sf v}^{(q)} = 1$. Then ${\sf u}^{(q)\dagger}{\sf v}^{(q')} = \delta_{qq'}$
 and ${\sf U}^\dagger\,{\sf V} = {\sf V}\,{\sf U}^\dagger = {\sf I}_{\rm Fl}$ \cite{Wilkinson}. Therefore
\begin{equation}
    {\sf U}^\dagger\left[(\omega^{(j)} + \omega_{\rm p})\,{\sf I}_{\rm Fl} - {\sf F}/\hbar\right]{\sf V} = {\sf \Lambda}^{(j)}
\end{equation}
where ${\sf \Lambda}^{(j)}$ is the diagonal matrix of elements
$(\omega^{(j)} + \omega_{\rm p} - \epsilon^{(q)}/\hbar) \,\delta_{qq'}$. It follows from equation~(\ref{eq:Feq}) that
\begin{equation}
{\sf U}^\dagger\left[(\omega^{(j)} + \omega_{\rm p})\,{\sf I}_{\rm Fl} - {\sf F}/\hbar\right]{\sf V}\,{\sf U}^\dagger\,{\sf x}^{(j)} = {\sf U}^\dagger\,{\sf b}^{(j)},
\end{equation}
and therefore\footnote{This result can also be derived directly from the closure relation \cite{Wilkinson} $\sum_q {\sf v}^{(q)}{\sf u}^{(q)\dagger} = {\sf I}_{\rm Fl}$ and from the spectral decomposition of the operator $(\omega^{(j)} + \omega_{\rm p})\,{\sf I}_{\rm Fl} - {\sf F}/\hbar$, i.e., $(\omega^{(j)} + \omega_{\rm p})\,{\sf I}_{\rm Fl} - {\sf F}/\hbar = \sum_q (\omega^{(j)} + \omega_{\rm p} - \epsilon^{(q)}/\hbar){\sf v}^{(q)}{\sf u}^{(q)\dagger}$. It has a similar mathematical structure as expressions of the ionisation rates derived from the spectral decomposition of the resolvent operator for atoms interacting with a weak probe beam in the presence of a strong static electric field \cite{Alvarez1989,Alvarez1989b} or a strong laser field \cite{Madajczyk1992,Jaron2000}.}
\begin{align}
    {\sf x}^{(j)} &= {\sf V}\,[{\sf \Lambda}^{(j)}]^{-1}\,{\sf U}^\dagger\,{\sf b}^{(j)}
    = \sum_q \frac{{\sf u}^{(q)\dagger}{\sf b}^{(j)}}{\omega^{(j)}+\omega_{\rm p} - \epsilon^{(q)}/\hbar}\,{\sf v}^{(q)}.
\label{eq:xexpanded}
\end{align} 
More explicitly,
\begin{align}
    \rho_{ij;N}^{\rm wp} &= 
     \sum_q \frac{{\sf u}^{(q)\dagger}{\sf b}^{(j)}}{\omega^{(j)}+\omega_{\rm p} - \epsilon^{(q)}/\hbar }\,\rho_{ij;N}^{(q)},
\label{eq:xexpanded2}
\end{align}
where $\rho_{ij;N}^{(q)}$ is the component of the vector ${\sf v}^{(q)}$ corresponding to the coherence $\rho_{ij;N}^{\rm wp}$.
Calculating the coherences in this way is completely equivalent to solving equation~(\ref{eq:steadysystem}). However, this alternative formulation makes it possible to identify the contribution of the different dressed states to the susceptibility $\chi(\omega_{\rm p},\omega_{\rm c},{\cal E}_{\rm c})$. It is also advantageous in calculations of $\chi(\omega_{\rm p},\omega_{\rm c},{\cal E}_{\rm c})$ over a range of probe frequencies since the  eigenvalues $\epsilon^{(q)}$, the eigenvectors ${\sf v}^{(q)}$ and the inner products ${\sf u}^{(q)\dagger}{\sf b}^{(j)}$ do not depend on $\omega_{\rm p}$ and can be calculated once and for all. In particular, this result makes it easier to average the susceptibility over a statistical distribution of values of $\omega_{\rm p}$, as may be required, e.g., for taking inhomogeneous broadening into account \cite{Potvliege2024}.

Finally, we note that these results must be corrected if the coherences $\rho_{ij}$ also decay not only through energy relaxation but also through some pure dephasing mechanisms such as collisional broadening or phase fluctuation of the probe field. Denoting the corresponding additional dephasing rates by $\gamma_{ij}$,
$\Gamma^{(i)}/2$ should then be replaced by $\Gamma^{(i)}/2+\gamma_{ij}$ in equations (\ref{eq:drhowp}), (\ref{eq:drhowpharmonic}), (\ref{eq:steadysystem}) and (\ref{eq:zerocfield}), and  $\omega^{(j)}+\omega_{\rm p} - \epsilon^{(q)}/\hbar$ should be replaced by $\omega^{(j)}+\omega_{\rm p} - \epsilon^{(q)}/\hbar + i\gamma_{ij}$ equations (\ref{eq:xexpanded}) and (\ref{eq:xexpanded2}).

\section{Numerical illustration}
\label{section:numerical}

\begin{figure}
\begin{center}
\includegraphics[width=0.9\textwidth]{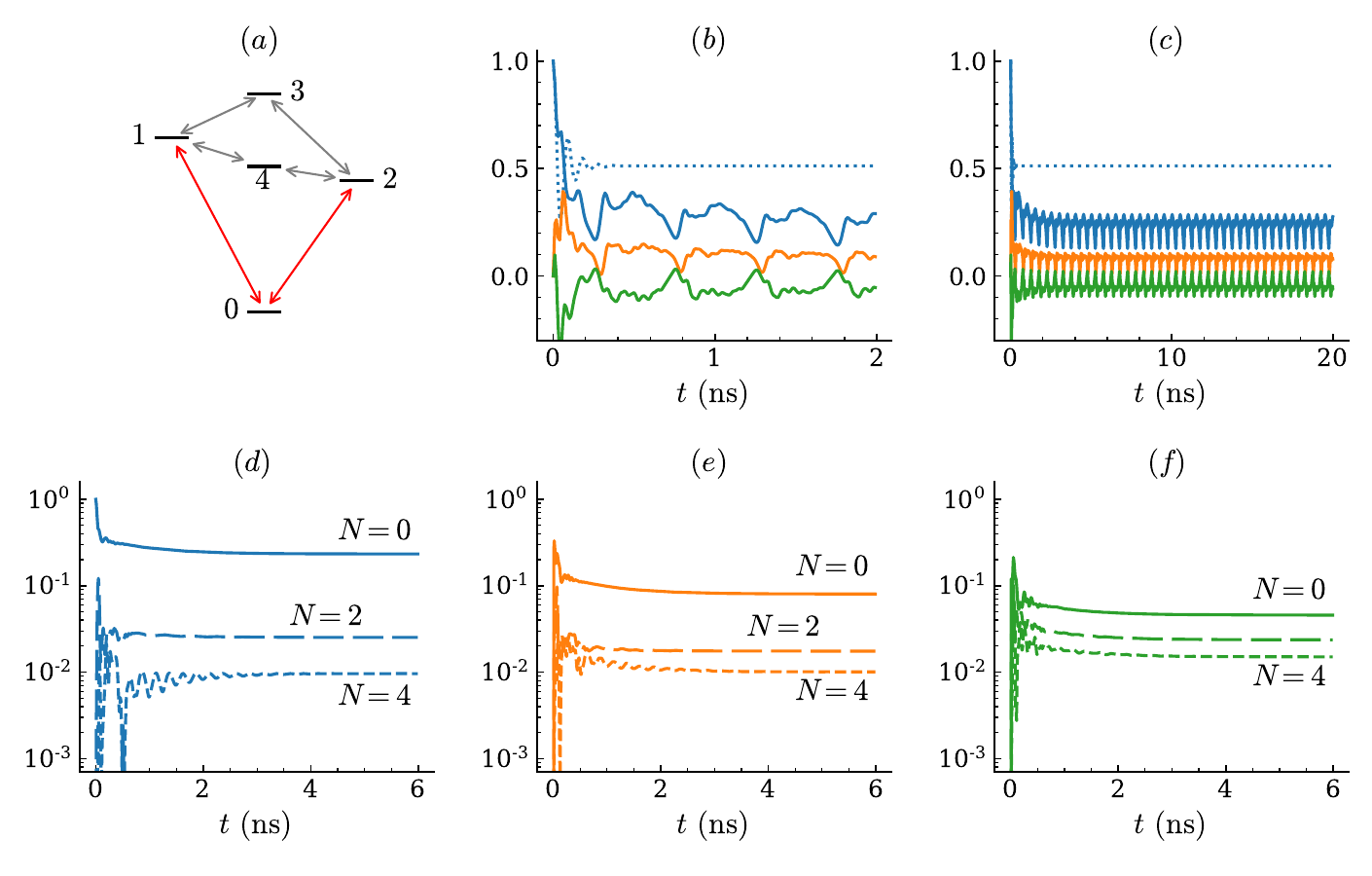}
\end{center}
\caption{($a$) The five-state system considered in Section~\ref{section:numerical}: states~1 and 2 are addressed from state~0 by the probe field {\color{black} and} coupled to states 3 and 4 by the coupling field. ($b$) and ($c$) Dotted blue curves: time evolution of $\rho_{00}^{\rm rw}(t)$ in the absence of the coupling field. Solid curves, from top to bottom: time evolution of $\rho_{00}^{\rm rw}(t)$ (solid blue curves), $\mbox{Im}\,\rho_{10}^{\rm rw}(t)$ (solid orange curves) and $\mbox{Im}\,\rho_{20}^{\rm rw}(t)$ (solid green curves) in the presence of the coupling field. ($d$) Time evolution of $|\mbox{Re}\,\rho_{00;N}^{\rm rw}(t)|$ for $N=0$ (solid curve), $N=2$ (long-dashed curve) and $N=4$ (short-dashed curve). ($e$) The same as ($d$) for $|\mbox{Im}\,\rho_{10;N}^{\rm rw}(t)|$. ($f$) The same as ($d$) for $|\mbox{Im}\,\rho_{20;N}^{\rm rw}(t)|$.}
\label{fig:figure1}
\end{figure}

\begin{figure}
\begin{center}
\includegraphics[width=0.9\textwidth]{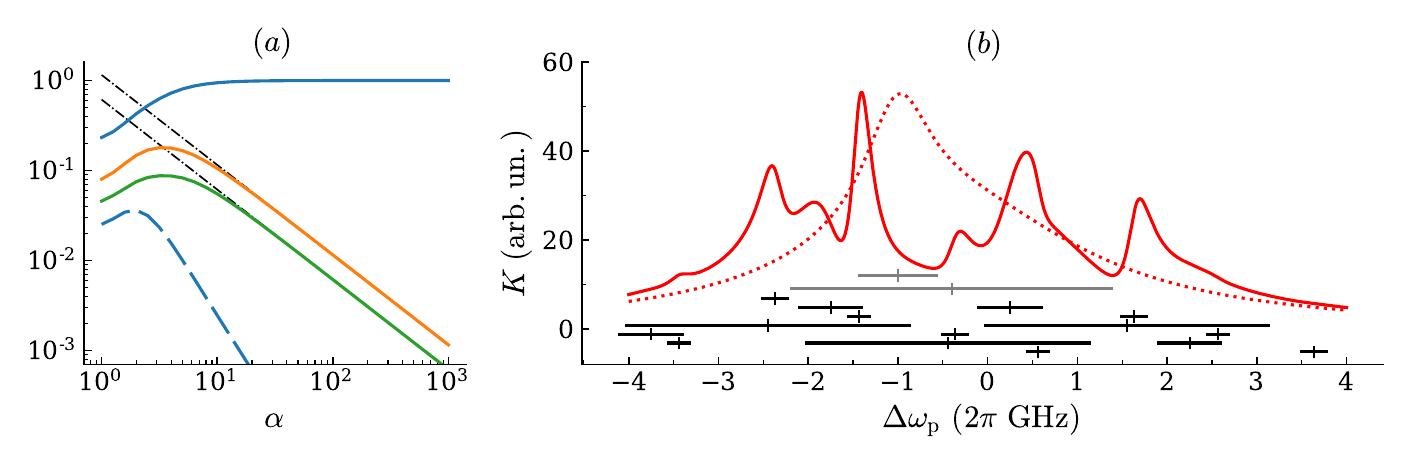}
\end{center}
\caption{($a$) From top to bottom, variation with the scaling parameter $\alpha$ of $\rho_{00;0}^{\rm rw}(t)$ (solid blue curve), of $|\mbox{Im}\,\rho_{10;0}^{\rm rw}(t)|$ (solid orange curve), of $|\mbox{Im}\,\rho_{20;0}^{\rm rw}(t)|$ (solid green curve), and of $|\mbox{Re}\,\rho_{00;2}^{\rm rw}(t)|$ (dashed blue curve), at $t = 200$~ns. The dash-dotted black lines show how {\color{black} the weak probe} $|\mbox{Im}\,\rho_{10;0}^{\rm wp}(t)|$ and $|\mbox{Im}\,\rho_{20;0}^{\rm wp}(t)|$ vary for the same parameters.
($b$) The variation of the absorption coefficient with probe frequency in the presence of the coupling field (solid red curve and black markers) and in the absence of the coupling field (dotted red curve and grey markers). The markers refer to the energy and width of the dressed states contributing most to the absorption spectrum in the range of frequencies spanned by the figure: the real part of the quasienergy of each of these dressed states is identified by a {\color{black} vertical tick mark} located at the corresponding value of $\Delta\omega_{\rm p}^{(q)}$ and its decay width is indicated by a horizontal bar (see text).
}
\label{fig:figure2}
\end{figure}

{\color{black} We now illustrate the theory developed in the previous section by numerical calculations for a simple toy model.} An application to a much more complicated case is reported elsewhere \cite{Brewin2024}. {\color{black} The system we are considering here} comprises only five states, namely a single lower energy state (state~0, here the only element of group~$A$) and four higher energy states (states 1, 2, 3 and 4, forming group~$B$). As shown in figure~\ref{fig:figure1}$a$, state~0 is coupled to states 1 and 2 by the probe field while states 1 and 2 are coupled to states 3 and 4 by the coupling field. States 1 and 2 decay to state~0 only, whereas states 3 and 4 are assumed to have zero decay rates. The parameters of the model are largely arbitrary and do not correspond to any actual system. We assume a frequency of 1~GHz for the coupling field and, in most of the calculations, a frequency of 100~THz for the probe field (this frequency is varied around 100~THz in figure~\ref{fig:figure2}b).
State~0 being the only member of group~$A$, we choose $\omega^{\rm ref}$ to coincide with $\omega^{(0)}$ so that $\delta\omega^{(0)} = 0$. We set the detunings $\Delta_{\rm p}^{(i)}$ and the Rabi frequencies $\Omega_{{\rm c};ij}$ to the values specified in table~\ref{table:table1}. We also assume that $\langle 0 |
 \hat{D}_z | 1 \rangle = 2\,\langle 0 |
 \hat{D}_z | 2 \rangle$. For consistency, we thus set $\Omega_{{\rm p};10} = 2 \,\Omega_{{\rm p};20}$ and $\Gamma^{(1)} = 4 \,\Gamma^{(2)}$. Specifically, we take $\Gamma_1= 2\pi\times3.6~{\rm GHz}$, $\Gamma_2 = 2\pi\times 0.9~{\rm GHz}$ and, when obtaining the results shown in figure~\ref{fig:figure1}, $\Omega_{{\rm p};10} = 2\pi\times 10~{\rm GHz}$ and $\Omega_{{\rm p};20} = 2\pi\times 5~{\rm GHz}$ (other values of $\Omega_{{\rm p};10}$ and  $\Omega_{{\rm p};20}$ are used in figure~\ref{fig:figure2}). 
 \begin{table}
 \caption{Detunings and Rabi frequencies used in the examples, in units of $2\pi$~GHz.}
 \label{table:table1}
 \begin{center}
 \begin{tabular}{lcccccccc}
 Examples &
 $\Delta_{\rm p}^{(1)}$ &
 $\Delta_{\rm p}^{(2)}$ &
 $\Delta_{\rm p}^{(3)}$ &
 $\Delta_{\rm p}^{(4)}$ &
 $\Omega_{{\rm c};13}$ &
 $\Omega_{{\rm c};14}$ &
 $\Omega_{{\rm c};23}$ &
 $\Omega_{{\rm c};24}$ \\[2mm]
 \hline \\[-3mm]
Figures~\ref{fig:figure1} and \ref{fig:figure2}$a$ &
0.4 & 1.0 & $-0.2$ & 0.8 &
9 & 11 & 6 & 9
\\
Figure~\ref{fig:figure2}$b$ &
\multicolumn{4}{c}{variable} &
0.9 & 1.1 & 0.6 & 0.9
 \end{tabular}
 \end{center}
 \end{table}

\begin{table}
{\color{black}
 \caption{Convergence of the solutions of Eq.~(\ref{eq:drndtmatrix}) to the solutions of Eq.~(\ref{eq:Lindblad3}) for increasingly wide ranges of the values of $N$ included in the calculation, for the system considered in Figure~\ref{fig:figure1}. The relevant detunings and Rabi frequencies are given in table~\ref{table:table1}.}
 \label{table:table2}
 \begin{center}
 \begin{tabular}{cccc}
 \parbox{3.2cm}{\centering $[N_{\rm min},N_{\rm max}]$} &
  \parbox{3.2cm}{\centering $\rho_{00}^{\rm rw}(t = 20~\mu\mbox{s})$} &
  \parbox{3.2cm}{\centering $\mbox{Im}\,\rho_{10}^{\rm rw}(t = 20~\mu\mbox{s})$} &
  \parbox{3.2cm}{\centering $\mbox{Im}\,\rho_{20}^{\rm rw}(t = 20~\mu\mbox{s})$} \\[2mm]
 \hline{} \\[-3mm]
$[-10,10]$ &$0.2774771$ & $0.0761685$ & $-0.0615402$ \\
$[-20,20]$ &$0.2759140$ & $0.0797740$ & $-0.0564471$ \\
$[-30,30]$ &$0.2760659$ & $0.0794110$ & $-0.0568168$ \\
$[-40,40]$ &$0.2753107$ & $0.0794474$ & $-0.0564466$ \\
$[-50,50]$ &$0.2753153$ & $0.0793790$ & $-0.0565258$ \\[1mm]
 Eq.~(\ref{eq:Lindblad3}) &$0.2753153$  & $0.0793790$   & $-0.0565258$ 
 \end{tabular}
 \end{center}
 }
 \end{table}

How some of the populations and coherences vary in time for this set of parameters is shown in figure~\ref{fig:figure1}$b$--$f$: The time-dependent harmonic components $\rho_{ij;N}^{\rm rw}(t)$ were calculated by numerically integrating equation~(\ref{eq:drndtmatrix}) and the density matrix by summing these harmonic components as per equation~(\ref{eq:rhoijexp}).
{\color{black} The range of values of $N$ used in the computation was wide enough that the results were not significantly affected by the truncation of the system~(\ref{eq:drndt}) to a finite number of equations: we set $N_{\rm min} = -50$ and $N_{\rm max} = 50$. As indicated by table~\ref{table:table2}, this choice ensured that the populations and coherences calculated from equations~(\ref{eq:rhoijexp}) and (\ref{eq:drndtmatrix}) matched those calculated from equation~(\ref{eq:Lindblad3}) to seven decimal places (a smaller range of values of $N$ would be sufficient for weaker fields). Results calculated are also presented for the case of a zero coupling field (the dotted curves in figures~\ref{fig:figure1}$a$ and \ref{fig:figure1}$b$).
}

How the ground state population $\rho_{00}^{\rm rw}(t)$ and the imaginary parts of the coherences $\rho_{10}^{\rm rw}(t)$ and $\rho_{20}^{\rm rw}(t)$ vary immediately after the probe laser is turned on is shown in figure~\ref{fig:figure1}$b$ and, over a longer time period, in figure~\ref{fig:figure1}$c$. In the absence of the coupling field (i.e., for $\Omega_{{\rm c};ij} \equiv 0$), the populations and coherences rapidly settle to constant values. However, they continue to oscillate markedly past the initial transients when the coupling field is strong. As shown by the solid curves, they settle into a state of periodic oscillation of period $4\pi/\omega$ when $t \rightarrow \infty$. By contrast, the harmonic components converge to constant values, reached asymptotically after a short phase of damped oscillations (figures \ref{fig:figure1}$d$--$f$, no results are shown for $N = 1$ and 3 because $\rho_{00;N}^{\rm rw}(t)$, $\rho_{10;N}^{\rm rw}(t)$ and $\rho_{20;N}^{\rm rw}(t)$ are identically zero for odd values of $N$). One can also note the importance of the $N \not= 0$ harmonic components of the ground state population although {\color{black} this state} does not directly interact with the coupling field in the present model. These harmonic components are responsible for the oscillation of $\rho_{00}^{\rm rw}(t)$. {\color{black} They} arise from the indirect interaction of the ground state with the coupling field originating from its coupling to states 1 and 2 by the probe field.

The results shown in figures~\ref{fig:figure1}$b$--$f$ refer to a relatively strong probe field. Convergence to the predictions of the weak probe approximation for decreasing values of ${\cal E}_{\rm p}$ is illustrated by figure~\ref{fig:figure2}$a$, which shows results obtained by integrating equation~(\ref{eq:drndtmatrix}) for values of $\Omega_{{\rm p};10}$ and $\Omega_{{\rm p};20}$ scaled by a factor $1/\alpha$: the probe field has the same strength as in figures~\ref{fig:figure1}$b$--$f$ when $\alpha=1$, but is weaker by a factor $1/\alpha$ when $\alpha > 1$. The calculation is done for a sufficiently large value of $t$ (200~ns) that the harmonic components $\rho_{ij;N}^{\rm rw}(t)$ have become constant in time. As expected, and as can be seen from the figure, $\rho_{00;N=0}^{\rm rw}(t)$ tends to 1 and $\rho_{00;N\not= 0}^{\rm rw}(t)$ tends to 0 when $\alpha$ increases. Moreover, $\rho_{10;0}^{\rm rw}(t)$ and $\rho_{20;0}^{\rm rw}(t)$ tend to the values predicted by equation~(\ref{eq:steadysystem}), namely $\rho_{10;0}^{\rm wp}$ and $\rho_{20;0}^{\rm wp}$. These values are proportional to ${\cal E}_p$, thus proportional to $1/\alpha$ here. (As can be deduced from a perturbative analysis and is apparent from the figure, $\rho_{00;2}^{\rm rw}(t) \propto 1/\alpha^2$.) 

Finally, part ($b$) of figure~\ref{fig:figure2} illustrates the use of this formalism in the calculation and analysis of an absorption spectrum. The weak probe approximation is assumed to hold. As in part ($a$) of the figure, the system is also taken be in the stationary state it relaxes into {\color{black} after the probe field has been turned on, past the initial transients.} Here, however, $\omega_{\rm p}$ is varied: we set 
\begin{equation}
\omega_{\rm p} = \omega_{{\rm p}0}+\Delta\omega_{\rm p},
\end{equation}
with $\omega_{{\rm p}0}/(2\pi) =  100$~THz. The detunings $\Delta_{\rm p}^{(i)}$ thus vary with $\Delta\omega_{\rm p}$ and have the same values as in Figures~\ref{fig:figure1} and \ref{fig:figure2}$a$ when $\Delta\omega_{\rm p} = 0$. The susceptibility and the absorption coefficient are calculated as functions of $\Delta\omega_{\rm p}$, by way of equations~(\ref{eq:chiwp}) and (\ref{eq:Kcoef}) respectively, with the products $N_{\rm d}\langle 0 | \hat{D}_z | 1 \rangle / (\epsilon_0{\cal E}_{\rm p})$ and $N_{\rm d}\langle 0 | \hat{D}_z | 2 \rangle / (\epsilon_0{\cal E}_{\rm p})$ set to values sufficiently low that $|\chi(\omega_{\rm p},\omega_{\rm c},{\cal E}_{\rm c}| \ll 1$. 

The resulting absorption spectrum is plotted in figure~\ref{fig:figure2}$b$, both for the values of $\Omega_{{\rm c};ij}$ stated in table~\ref{table:table1} (the solid curve) and for the case of a zero coupling field ($\Omega_{{\rm c};ij}\equiv 0$, the dotted curve). The relevant quasienergies are also indicated in the figure. {\color{black} As shown by equation~(\ref{eq:xexpanded}), a dressed state of quasienergy $\epsilon^{(q)}$ contributes to the coherences $\rho_{i0}^{\rm wp}$ mostly for values of $\Delta\omega_{\rm p}$ in the range
$$\Delta\omega_{\rm p}^{(q)}-|\mbox{Im}\,\epsilon^{(q)}/\hbar| \leq \Delta\omega_{\rm p} \leq \Delta\omega_{\rm p}^{(q)}+|\mbox{Im}\,\epsilon^{(q)}/\hbar|
$$ where $\Delta\omega_{\rm p}^{(q)}$ is the value of $\Delta\omega_{\rm p}$ such that $\mbox{Re}\,\epsilon^{(q)}/\hbar + \omega^{(0)} + \omega_{\rm p} = 0$.} Accordingly, we identify the dressed states contributing most to the spectrum by {\color{black} tick marks} positioned at the corresponding values of $\Delta\omega_{\rm p}^{(q)}$ and we represent their widths by bars of length $2\,|\mbox{Im}\,\epsilon^{(q)}/\hbar|$. (The vertical position of these markers is arbitrary and has no physical meaning.) 

The ground state interacts only with the bare states 1 and 2 when ${\cal E}_{\rm c}=0$. Diagonalising the matrix ${\sf F}$, in that case, yields quasienergies either of the form $N\hbar\omega_{\rm c}-\hbar(\omega^{(1)} - i\Gamma^{(1)}/2)$ or of the form $N\hbar\omega_{\rm c}-\hbar(\omega^{(2)} - i\Gamma^{(2)}/2)$, $N = 0,\pm 1, \pm 2$,\ldots Only the two solutions for which $\epsilon^{(q)}$ is either exactly $-\hbar(\omega^{(1)} - i\Gamma^{(1)}/2)$ or $-\hbar(\omega^{(2)} - i\Gamma^{(2)}/2)$ are relevant here, because, in the notation of equation~(\ref{eq:xexpanded}), ${\sf u}^{(q)}{\sf b}^{(0)} = 0$ for the other solutions in the absence of the coupling field. The corresponding states overlap in energy within their respective decay width, as shown by the grey horizontal bars in the figure. The overlap results in a single non-Lorentzian peak in the absorption spectrum.  
This peak splits into multiple structures when the coupling field is on, for the relatively large values of $\Omega_{{\rm c};ij}$ assumed in the calculation. The structures originate from the interfering contributions of many dressed states. For clarity, only the 15 states contributing most to the spectrum in the range of frequencies considered are identified in the figure.

{\color{black}
\section{Conclusions}
\label{section:conclusions}

To conclude, we have shown how the familiar theory of absorption of a weak probe laser beam by a linear medium generalises to the case where the medium is also addressed by a strong oscillating field coupling states with a significant decay width. The Floquet formalism is the natural framework for modelling such systems. Calculating the absorption spectrum is particularly simple for weak probe beams, as in the weak field limit the relevant steady-state coherences can be obtained by solving a system of linear equations of reduced dimensionality, equation~(\ref{eq:steadysystem}). Taking decay mechanisms in this case simply amounts to adding an imaginary part to the state energies, with no need of solving the Lindblad master equation in its full complexity; however, these two formulations are mathematically equivalent. Moreover, the Floquet approach makes it possible to analyse the absorption spectrum in terms of contributions of dressed excited states spawned by the coupling field, in the same way as it can be analysed in terms of contributions of bare excited states in the absence of the coupling field.

The methods described in this article have already been applied to the study of broad excitons coupled to each other by a strong microwave field~\cite{Brewin2024}. They are generally applicable to any system driven by a strong oscillating field while probed by a weak field, as long as the rotating wave approximation and the Beer-Lambert law apply for the latter.

}
\vskip2pc

\section*{Acknowledgements}

This work has much benefited from discussions with Robin Shakeshaft, a number of years ago, in regard to the non-Hermitian Floquet theory of multiphoton processes in strong laser fields. It has also benefited from discussions with the late Howard Taylor in regard to the related calculations reported in reference~\cite{Madajczyk1992}, 
and with Alistair Brewin and Matt Jones in regard to the use of Floquet calculations in the spectroscopy of Rydberg excitons.

\end{document}